**DYNAMICS OF INTERPLANETARY PLANETESIMAL RINGS.** Bruce D. Lindsay and Truell W. Hyde, CASPER (Center for Astrophysics, Space Physics and Engineering Research), P. O. Box 97316, Baylor University, 76798-7316, USA; e-mail: Truell_Hyde@baylor.edu.

**Introduction**: Among the problems yet to be solved by current theories of solar system formation is the origin of the Jupiter–family comets. These comets are characterized by relatively short orbital periods as compared to comets originating from the Oort cloud. Most are thought to come from the trans-Neptunian region known as the Kuiper Belt, a region filled with planetesimals that have remained almost undisturbed since the birth of the solar system. However, it is also possible that some of these objects could have come from thin but stable rings of planetesimals lying between the existing giant planets [1]. Another interesting open question is whether the recently suggested migration of protoplanets could be due to interactions with nearby planetesimal swarms. This possibility is bolstered by the discovery [2] that planetesimal swarms having mass densities larger than a critical value can trigger significant migration in nearby protoplanets. Such a process could provide a clue toward explaining the recent discovery of extrasolar planets orbiting at small distances from their parent stars.

**Simulation Model:** A fifth-order Runge-Kutta algorithm was implemented to examine the above theories. The model integrated the trajectories of both the planets and planetesimals as they orbited the Sun. All simulations begin with four planets (Jupiter, Saturn, Uranus, and Neptune) and 496 planetesimals, resulting in a total of 500 bodies (excluding the Sun) being tracked. The initial positions and velocities of the giant planets are shown in Table 1 and were determined using orbital parameters listed in previously published works [3]. The planetesimals are placed in circular, coplanar orbits with their semimajor axes uniformly scattered over a range between 20 and 30 AU from the Sun, and their velocities initially defined to be Keplerian. Each simulation collects data over a 20,000 year period. Unlike many simulations in the literature, the effects of the mutual gravitational forces between all the bodies in the system were taken into account. This allows for an explicit examination of the reaction of the giant planets to the presence of the smaller bodies. A complete investigation of such interparticle forces is necessary to determine how such planetesimal rings might affect the stability of the system as a whole.

**Results and Conclusions:** Fig. 1 shows the final state for planetesimals remaining within the ring at the end of the simulation, revealing their final values for semimajor axes and eccentricities (which are shown on a natural logarithmic scale). One easily distinguishable feature is the apparent depletion of planetesimals in the 20-21 AU and 28-29 AU regions. These gaps are most likely explained by the presence of Uranus and Neptune since mean motion resonances associated with these planets can increase the eccentricities of the planetesimals, eventually causing them to be scattered out of the ring. Fig. 1 also shows a peak in the 25-26 AU range, which is consistent with Holman's [1] identification of an enhanced stability region between 24 and 27 AU. Interestingly, Fig. 1 also shows that the majority of planetesimals having low final eccentricities ($< 0.01$) are found in this region as well. As a result, these planetesimals would be those most likely to produce a stable ring for any extended period of time. Taken together, the above seems to imply that a planetesimal ring could exist in this region.

If such a stable ring existed, it might provide the critical mass density needed to trigger the aforementioned migration process. Since (as mentioned) the majority of planetesimals in regions close to the semimajor axes of Uranus and Neptune are scattered out of the system, it is logical to assume these two planets would interact more strongly with such a planetesimal ring than would either Jupiter or Saturn. The data given in Table 2 verifies this assumption, showing that Neptune does indeed migrate inward as the planetesimal mass increases, with this motion reaching a maximum when the planetesimal masses exceed $10^{-3}$ Earth masses. This yields a ring mass of $4.96 \times 10^{-1}$ Earth masses, a figure which is in rough agreement with Holman's [1] observational maximum mass limit. As can be seen, Neptune only shows significant migration when the local planetesimal mass density exceeds 0.13 kg/m$^3$, the critical density derived from Holman's formula [1]. Any such motion ends at the gap between 28 and 29 AU. It can also be seen from the data that Jupiter, Saturn, and Uranus would be affected by scattered planetesimals as well. A slight inward drift by Jupiter and an outward drift by Saturn are the primary results of interactions with the planetesimals that are scattered into those parts of the Solar System.

Thus, the data suggests that planetesimal swarms might also (at least partially) explain protoplanetary migration.

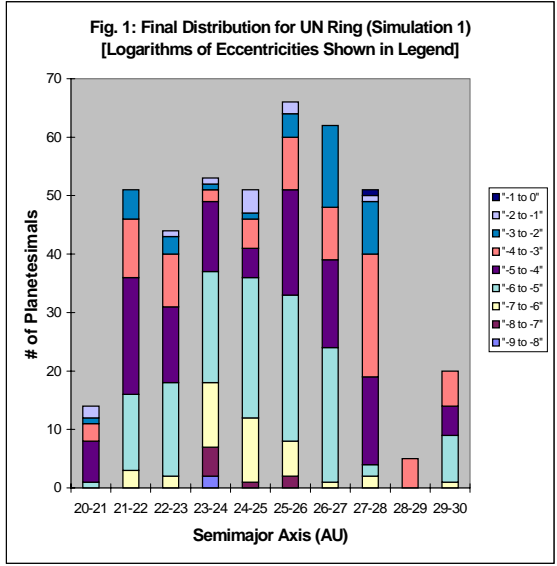

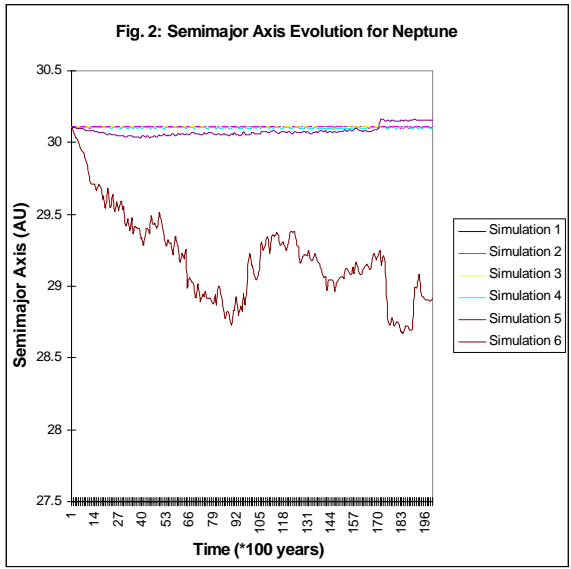

Table 1: Initial Conditions for Giant Planets

| Planet | Jupiter | Saturn | Uranus | Neptune |
|---|---|---|---|---|
| A (AU) | 5.207 | 9.553 | 19.219 | 30.111 |
| E | 0.04749 | 0.05274 | 0.04641 | 0.00820 |
| i (rad) | 0.02277 | 0.04338 | 0.01348 | 0.03089 |

Table 2: Collective Simulation Results
________________

| Simulation # | 1 | 2 | 3 |
|---|---|---|---|
| Log(Mp/Me) | -6 | -5 | -4 |
| Melted: | 2 | 0 | 0 |
| Escaped: | 2 | 5 | 1 |
| Scattered Inside: | 12 | 6 | 11 |
| Scattered Outside: | 63 | 67 | 67 |
| Jupiter Crossers: | 1 | 0 | 0 |
| Saturn Crossers: | 3 | 1 | 0 |
| Uranus Crossers: | 7 | 8 | 11 |
| Neptune Crossers: | 28 | 34 | 30 |
| Mean SA (Jupiter): | 5.1936 | 5.1936 | 5.1935 |
| (AU) | ± 0.0013 | 0.0014 | 0.0013 |
| Mean SA (Saturn): | 9.5357 | 9.5356 | 9.5357 |
| (AU) | ± 0.0058 | 0.0059 | 0.0058 |
| Mean SA (Uranus): | 19.1852 | 19.1852 | 19.1804 |
| (AU) | ± 0.005 | 0.005 | 0.0041 |
| Mean SA (Neptune): | 30.1083 | 30.1085 | 30.1066 |
| (AU) | ± 0.0038 | 0.0038 | 0.002 |

________________

| Simulation # | 4 | 5 | 6 |
|---|---|---|---|
| Log(Mp/Me) | -3 | -2 | -1 |
| Melted: | 2 | 2 | 5 |
| Escaped: | 7 | 5 | 11 |
| Scattered Inside: | 10 | 13 | 34 |
| Scattered Outside: | 65 | 66 | 112 |
| Jupiter Crossers: | 1 | 0 | 2 |
| Saturn Crossers: | 2 | 4 | 7 |
| Uranus Crossers: | 11 | 17 | 40 |
| Neptune Crossers: | 43 | 40 | 96 |
| Mean SA (Jupiter): | 5.1935 | 5.1933 | 5.1931 |
| (AU) | ± 0.0013 | 0.0013 | 0.0013 |
| Mean SA (Saturn): | 9.5357 | 9.5356 | 9.5362 |
| (AU) | ± 0.0059 | 0.0059 | 0.0061 |
| Mean SA (Uranus): | 19.1784 | 19.0707 | 19.3106 |
| (AU) | ± 0.0036 | 0.0299 | 0.0797 |
| Mean SA (Neptune): | 30.1023 | 30.0788 | 29.2018 |
| (AU) | ± 0.0027 | 0.035 | 0.3087 |